\documentclass{emulateapj}

\usepackage{epsfig}
\tightenlines

\newcommand \lsim{\mathrel{\rlap{\lower4pt\hbox{\hskip1pt$\sim$}}
    \raise1pt\hbox{$<$}}}
\newcommand \gsim{\mathrel{\rlap{\lower4pt\hbox{\hskip1pt$\sim$}}
    \raise1pt\hbox{$>$}}}

\newcommand     \gtsim  {\gtrsim}                

\newcommand     \ltsim  {\lesssim}               

\newcommand{\tff}{t_{\rm ff}}

\newcommand{\beq}{\begin{equation}}
\newcommand{\eeq}{\end{equation}}
\newcommand{\beqa}{\begin{eqnarray}}
\newcommand{\eeqa}{\end{eqnarray}}

\newcommand{\krho}{{k_\rho}}

\newcommand{\rcl}{R_{\rm cl}}

\newlength{\figwidth}
\countdef\decade=200
\advance\decade by \year
\countdef\hours=201
\advance\hours by \time
\divide\hours by 60
\countdef\mins=202
\advance\mins by \hours
\multiply\mins by 60
\multiply\hours by 100
\countdef\miltime=203
\advance\miltime by \hours
\advance\miltime by \time
\advance\miltime by -\mins

\begin{document}

\title{Equilibrium Star Cluster Formation}


\author{Jonathan C. Tan}
\affil{Dept. of Astronomy, University of Florida, Gainesville, FL 32611, USA\\
and Inst. of Astronomy, Dept. of Physics, ETH Z\"urich, 8093 Z\"urich, Switzerland}
\author{Mark R. Krumholz}
\affil{Hubble Fellow, Department of Astrophysics, Princeton University, Princeton, NJ 08544, USA}
\author{Christopher F. McKee}
\affil{Depts. of Physics \& Astronomy, UC Berkeley, Berkeley, CA 94720, USA}

\begin{abstract}
  We argue that rich star clusters take at least several local
  dynamical times to form, and so are quasi-equilibrium structures
  during their assembly. Observations supporting this conclusion
  include morphologies of star-forming clumps, momentum flux of
  protostellar outflows from forming clusters, age spreads of stars in
  the Orion Nebula Cluster (ONC) and other clusters, and the age
  of a dynamical ejection event from the ONC. We show that
  these long formation timescales are consistent with the expected
  star formation rate in turbulent gas, as recently evaluated
  by Krumholz \& McKee. Finally, we discuss the implications of these
  timescales for star formation efficiencies, the disruption of gas by
  stellar feedback, mass segregation of stars, and the longevity of
  turbulence in molecular clumps.
\end{abstract}


\section{Introduction}\label{S:intro}

Star clusters are born from the densest gas clumps in giant molecular
clouds and are likely to be responsible for the majority of stars ever
formed (Lada \& Lada 2003).  The timescale over which a clump
transforms into a cluster is a basic constraint for theoretical
models. If formation takes only $\sim 1$ dynamical time, then star
formation is a result of the global collapse of the clump that
internal sources of feedback are insufficient to impede. If it takes
several dynamical times, then the clump gas must reach an approximate
equilibrium, with self-gravity resisted by internal sources of
pressure. In this case, star formation is a local process within the
protocluster.  Runaway gravitational instability occurs on scales much
smaller than the overall cluster, involving only a small fraction of
the mass at any given time. The cluster formation timescale also
determines how much dynamical mass segregation occurs in the gas-rich
phase, which may account for the observed central concentration of
massive stars in young clusters (Hillenbrand \& Hartmann 1998).

Elmegreen (2000) has argued that, over a wide range of scales, star
formation occurs in $\sim 1$ crossing time. For star clusters the
observational evidence cited for short formation times is the age
spread of stars and substructure in their spatial distributions.  We
examine these arguments in detail in \S2 and show that the evidence
for rich clusters in fact points to a considerably slower formation
process.  Hartmann, Ballesteros-Paredes, \& Bergin (2001) have argued
for star formation in one crossing time for low mass, distributed
star-forming regions like Taurus. Tassis \& Mouschovias (2004) have
presented counter arguments. However, neither of these analyses apply
to individual rich star clusters, on which we focus.

We define here the important timescales for a protocluster gas clump
of mass $M$, radius $R$, 1D velocity dispersion $\sigma$, density
$\rho$, and column density $\Sigma$. The free-fall time is $t_{\rm
ff}=(3\pi/32G\rho)^{1/2}$, and the dynamical, or crossing, time is
$t_{\rm dyn}=R/\sigma$.  The virial parameter, $\alpha_{\rm vir}
\equiv 5 \sigma^2 R/(G M)$ (Bertoldi \& McKee 1992), defines the
relationship of these two; for $\alpha_{\rm vir}\sim 1$, $t_{\rm
ff}\approx 0.5 \,t_{\rm dyn}$. Clumps are centrally concentrated
(Mueller et al. 2002), so the free-fall and dynamical times vary
by location within them. We therefore define $t_{\rm ff}$ and $t_{\rm
dyn}$ as mass-weighted averages over the region containing 90\% of the
mass, although in practice this may be difficult to determine due to
confusion on the outskirts of clusters. Finally, we define the
formation time $t_{\rm form}$ as the time over which $90\%$ of the
stars in a cluster form. Note, this is $2.3$ times the
exponentiation time $t_0$ in the accelerating star formation model of
Palla \& Stahler (2000).

In \S2 we discuss observational evidence that rich clusters take at
least several dynamical times to form, drawing on various stages of
the formation process. In \S3 we use a theoretical estimate of the
star formation rate to show that long formation times are to be
expected. Finally, we discuss the implications of this result in \S4.


\section{Observational Evidence for Slow Formation}\label{S:observation}

\subsection{Morphologies of Gas Clumps}

Shirley et al. (2003) observed CS ($J=5\rightarrow4$) emission from
about 60 dense gas clumps containing high-mass star formation. Their
surface densities and masses are very similar to more revealed rich,
young star clusters, so they are likely to be similar objects at an
early stage of formation. Shirley et al. determined clump aspect
ratios, defined as the ratio of major and minor axes of the 20\% of
peak contour, which is well detected and resolved. The distribution of
aspect ratios peaks at $1.26\pm0.22$, consistent with most clumps
being circularly symmetric. This morphology contrasts with
simulations of rapid cluster formation in which one or two gas
filaments tend to dominate (e.g. Bate, Bonnell, \& Bromm
2003). Non-periodic simulations with stellar feedback that run for
longer times need to be performed to see how long it takes to establish
spherical morphologies. 
From the CS data it is not possible to tell if the clumps of gas
relaxed before star formation started or after. Studies of earlier
stages of star cluster formation, as traced by the Infrared Dark
Clouds (Egan et al. 1998), can address this.

\subsection{Morphologies of Embedded Stars}

Since molecular clumps are turbulent, stars should form in
substructures, which will dissolve as the stars orbit. Thus, the
amount of substructure can constrain the formation time.  The
dissipation time of a substructure depends on whether it is bound
(Scally \& Clarke 2002) and the nature of its velocity dispersion
(Goodwin \& Whitworth 2004). If unbound, it dissipates in a time $\sim
t_{\rm dyn}$. Assuming all stars form in substructures at constant
rate, this implies $t_{\rm form} \sim (M/M_{\rm sub}) t_{\rm dyn}$,
where $M_{\rm sub}$ is the mass in substructures. If a subcluster is
bound, rather than dissolving it will sink to the cluster center due
to dynamical friction. The sinking time from radius $r_{\rm form}$ is $t_{\rm
sink} = 0.68 (r_{\rm form}/R)^2 (\Lambda/\ln\Lambda) t_{\rm dyn}$ (Binney \&
Tremaine 1987), where $\Lambda$ is the cluster to
most-massive-subcluster mass ratio. This assumes the cluster density
varies as $r^{-2}$, but a different exponent would not substantially
change the result. Assuming a constant star formation rate, so the
last subcluster formed at least a time $t_{\rm form}/\Lambda$ ago, that all
stars form in subclusters, and that on average subclusters form at
$r_{\rm form}=R/2$, the formation time and the mass of the largest visible
subcluster are related by
$t_{\rm form} \leq 0.17 (2 r_{\rm form}/R)^2 (\Lambda^2/\ln \Lambda) t_{\rm dyn}$.
To improve upon these approximate analytic estimates requires
global numerical simulations of stars forming from turbulent gas
(e.g. Schmeja \& Klessen 2006).

Comparing to observations, we find that, contrary to Elmegreen (2000),
they are consistent with long formation times. The ONC has quite
smooth contours of projected stellar surface density with no
significant substructure (Hillenbrand \& Hartmann 1998), giving no
limit on $t_{\rm form}$ (see also Scally \& Clarke 2002). In IC~348,
there are 8 small subclusters, but these contain only $\sim 10-20$
stars each, roughly $20\%$ of the total 345 stars in the cluster (Lada
\& Lada 1995), giving $t_{\rm form} \sim 5 t_{\rm dyn}$ even if all
subclusters are unbound. The largest subcluster has 18 members, so
$\Lambda \approx 19$ and $t_{\rm form} \leq 21 t_{\rm dyn}$ if
subclusters are bound. Note, we have neglected the artificial
enhancement of apparent substructure by patchy extinction. If this
effect is significant, the true ages may be even larger. In no case do
we find rich clusters with all or most of their mass in substructures,
which would imply $t_{\rm form} \approx t_{\rm dyn}$.

\subsection{Momentum Flux of Protostellar Outflows}

Tan \& McKee (2002) estimated the star formation rate in 8 clusters
using the observed momentum flux of protostellar outflows. Outflows
are magnetocentrifugally driven from the star and inner accretion disk
(e.g. Shu et al. 2000), expelling a fraction $f_w$ of the mass flux
$\dot{m}_*$ reaching the star. Outflow models find that the ratio $f_p
\equiv f_w v_w/v_K$ is constant to within $\sim 30\%$ (Najita \& Shu
1994), where $v_w$ is the outflow velocity and $v_K$ is the Keplerian
velocity at the equatorial radius of the star. Thus, the total
momentum flux $\dot{p}_w = f_p \dot{m}_* v_K$ is determined by the
star formation rate and the evolution of protostellar radii, with only
a weak dependence on the latter. Tan \& McKee (2002) find that, for
clusters with a Salpeter IMF from $0.1-120$ $M_{\odot}$, $\dot{p}_w /
\dot{M}_* \simeq 87$ km s$^{-1}$, where $\dot{M}_*$ is the total
star formation rate. Loss of momentum in outflow-outflow
interactions would lower this estimate, but this effect is probably
small as outflow jets are well-collimated. Thus, observations of
$\dot{p}_w$ give a measurement of $\dot{M}_*$ and hence the cluster
formation timescale. The measurement of $\dot{p}_w$ is quite
uncertain, but the data suggest that it would take $\sim 3-5$
dynamical times to transform $\sim 30\%$ of the mass into stars (Tan
\& McKee 2002). Although this is currently one of the more uncertain
methods of estimating $t_{\rm form}$, its accuracy should
improve as models and observations of individual protostellar outflows
are refined.

\subsection{Age Spreads of Pre-Main-Sequence Stars}

The age spread of stars in a cluster is a direct measure of the
formation time. Ages can be difficult to determine because they
require fitting observed luminosities and temperatures to pre-main
sequence models that have quite large systematic uncertainties,
particularly for low (sub-solar) mass stars. The models also depend on
uncertain parameters such as the deuterium abundance, the accretion
rate, and accretion geometry (Stahler 1988; Palla \& Stahler 1999,
hereafter PS99; Hartmann 2003), which influence the position of the
``birthline'', where stars first appear on the HR
diagram. Fortunately, the initial stellar contraction is quite rapid,
so the birthline position mostly affects ages $\lesssim 1$~Myr.
Patchy extinction also introduces systematic errors into age
estimates, as do unresolved binaries, although for these one may make
approximate statistical corrections (PS99). Photometric variability is
another potential source of error, but recent observations by
Burningham et al. (2005) find that variability cannot mimic age
spreads of several Myr. In summary, age estimates for intermediate
mass ($M \sim M_\odot$), older (age $\gtsim 1$~Myr) stars are
reasonably robust.

In the ONC, PS99 estimate ages of 258 stars with masses
$0.4<m_*/M_\odot <6.0$ from the sample of Hillenbrand (1997).
PS99 found 82 stars aged 0-1~Myr, 57 aged
1-2~Myr, 34 aged 2-3~Myr, 17 aged 3-4~Myr, 8 aged 4-5~Myr, 8 aged
5-6~Myr, 8 aged 6-7~Myr, and 6 aged 7-10~Myr. Hartmann (2003) has
argued that the oldest ages ($\sim 10$~Myr) may be due to a problem of
foreground contamination. 
We conclude that a significant fraction of the stars are 3~Myr
old. This is a lower limit to $t_{\rm form}$ since star formation is
still continuing in the cluster, and because the sample is potentially
incomplete for the oldest low mass stars.  This result is broadly
consistent with ONC age determinations based on Li abundances in
pre-main-sequence stars (Palla et al. 2005), a few of which imply ages
as large as $\sim 10$~Myr. Elmegreen (2000) estimates a density of
$n_{\rm H}=1.2\times 10^5\:{\rm cm^{-3}}$ for the gas clump from which
the ONC formed, giving a dynamical time of $2.5\times
10^5$ yr and an age $\geq$ 12 crossing times. Note
that Elmegreen argued star
formation was rapid in the ONC, adopting an age spread of only
1~Myr.

Elmegreen's dynamical time is probably somewhat low; a better estimate
is $t_{\rm dyn} = 0.95 (\alpha_{\rm vir} G)^{-1/2}
(M/\Sigma^3)^{1/4}$. We use this rather than relying on a measured
velocity dispersion because, for star clusters, this is often hard to
determine due to incompleteness, confusion, and variation in the
velocity dispersion with location in the cluster. The mass and surface
density are easier to measure. For gas systems the typical virial
parameter is $\alpha_{\rm vir}\approx 1.3$ (McKee \& Tan 2003). For
stellar systems we adopt a King model, which implies $\alpha_{\rm vir}
\geq 2.0$ if we take $\sigma$ to be the dispersion of the Maxwellian
velocity distribution. For the ONC, Hillenbrand \& Hartmann (1998)
find $M=4600$ $M_{\odot}$, $\Sigma=0.12$ g cm$^{-2}$, giving $t_{\rm
dyn} = 7\times 10^5$ yr and an implied cluster age of 4 dynamical
times.

In addition to the ONC, which has the best observational data, we can
make estimates of the formation and dynamical times for only a few
other rich, forming star clusters. Palla \& Stahler (2000) find age
spreads of $t_{\rm form}\approx 2.3$ Myr for both $\rho$~Ophiuchi and
IC~348. For $\rho$~Oph, the central cluster is embedded in the L1688
dark cloud. The cluster has a radius of $\approx 1$ pc, and the gas
mass within this radius (which dominates the total mass) is roughly
$1500$ $M_{\odot}$ (Loren 1989). The inferred dynamical time is
$7.6\times 10^5$ yr, implying a formation time of 3 dynamical times --
in a system that is still gas-dominated. The central cluster in
IC~348, which contains roughly half the stars and for which Palla \&
Stahler make their age estimate, has a radius of 0.5 pc and contains
$\approx 200$ $M_{\odot}$ of stars (Lada \& Lada 1995). Stars probably
dominate the mass (Herbig 1998), so we infer a dynamical time of $6
\times 10^5$ yr, giving a formation time of 4 dynamical times.

It is more difficult to determine both formation and dynamical times
for other systems, so, contrary to Elmegreen (2000), few conclusions
can be drawn. For example, Forbes (1996) did not find evidence for an age
spread in NGC~6531, but the analysis was insensitive to timescales
shorter than $\gtsim 3$~Myr. Hodapp \& Deane (1993) determined ages
up to 6 Myr for stars in L1641, but studied only 12 objects, did not
correct for binarity, and used relatively old pre-main sequence
tracks. Palla \& Stahler (2000) found formation times for
Taurus-Auriga, Lupus, Chamaeleon, Upper Scorpius, and NGC~2264 that
are $\gtrsim 3$~Myr, but the first three of these are not rich
clusters, Upper Scorpius has undergone too much dynamical spreading to
reliably estimate its dynamical time at formation, and NGC~2264 is too
distant for a complete census to reliably determine its mass
and column density, and thus its dynamical time. In NGC~3603,
Eisenhauer et al. (1998) find both young stars ($\ltsim 0.5$~Myr) and
Wolf-Rayet stars $2-3$~Myr old, but as with NGC~2264 it is too distant
to reliably determine a dynamical time.
Elmegreen (2000) comments that star clusters with age spreads $\sim
10$~Myr, e.g. NGC~1850, NGC~2004, NGC 4755, NGC~6611 and the Pleiades,
could be the result of ``multiple and independent star formation
events'', which then form a single cluster by merging or only appear
to be a single cluster because of projection. We would argue that in
the former case, the age spread is a true indication of the cluster
formation time, although, as Elmegreen points out, the relevant
dynamical timescale may need to be evaluated over a larger region that
hosted the initial subclusters. However, in all these systems, there
is no evidence to favor merging of independent subclusters over
continuous formation {\it in situ}.

\subsection{Age of a Dynamical Ejection Event in Orion}

Dynamical ejection events provide another method of age
estimation. One such event involving 4 massive stars (a binary and two
singles) that appear to have come from the ONC has been dated to $\sim
2.5$ Myr ago (Hoogerwerf, de Bruijne \& de Zeeuw 2001). The central
value of the time since this ejection event is $2.3\pm0.2$~Myr in
Hoogerwerf et al.'s analysis; however, if the cluster's distance of
about 450~pc is adopted, then the best estimate is 2.5~Myr. The
identification with the ONC is based upon the extrapolation of the
motion of the center of mass of the four stars from the ejection event
to the present day, leading to a predicted position coincident with the
ONC (uncertainties are a couple of pc).  This result implies that
2.5~Myr ago the ONC was already a rich cluster containing at least
four stars of spectral type earlier than O9/B0. Before this the stars
had to form and have enough time to find and eject each other in a
close interaction. Thus the estimate of 2.5~Myr is again a lower limit
to $t_{\rm form}$ for the ONC, so that $t_{\rm form}\geq 4 t_{\rm dyn}$.


\section{Theoretical Formation Timescale}\label{S:theory}

Krumholz \& McKee (2005, hereafter KM05) estimate the star
formation rate in supersonically turbulent gas, and we use this
result to compute how long star formation in a clump must
continue to reach $\sim 30\%$ efficiency.  Consider a clump with
density and pressure profiles $\rho\propto r^{-\krho}$
and $P\propto r^{-k_P}$. Hydrostatic equilibrium requires
that its mass, radius, and effective sound speed $c\equiv
(P/\rho)^{1/2}$ be related by $M=k_P c^2 R/G$. The effective sound
speed is related to the 1D velocity dispersion $\sigma_{\rm cl}$ by
$c=\phi_B^{1/2} \sigma_{\rm cl}$, where $\phi_B$ is a factor
accounting for magnetic support. The
clump velocity dispersion in terms of clump mass and surface
density is thus
\begin{equation}
\sigma_{\rm cl} = \left(\frac{G^2 \pi M \Sigma}{k_P^2 \phi_B^2}\right)^{1/4}
\rightarrow 
2.4\, M_3^{1/4} \Sigma_0^{1/4} \mbox{ km s}^{-1},
\end{equation}
where $M_3$ is the clump mass in units of $10^3$ $M_{\odot}$,
$\Sigma_0$ is column density in g cm$^{-2}$, and the numerical
evaluation is for the fiducial parameters of McKee \& Tan (2003)
($k_P=1$, $\phi_B=2.8$, and $\alpha_{\rm vir}=1.3$). Based on analysis
of the structure of turbulent media and comparison to numerical
simulations, KM05 find that the fraction of the mass forming stars
every free-fall time is $\mbox{SFR}_{\rm ff} \approx 0.073\,
\alpha_{\rm vir}^{-0.68} \mathcal{M}^{-0.32}$, where
$\mathcal{M}\equiv \sigma_{\rm cl}/c_s$ is the Mach number, and $c_s$
is the thermal sound speed. For our clump, this gives $\mbox{SFR}_{\rm
ff-cl} \approx 0.027 \,T_1^{0.16} (M_3 \Sigma_0)^{-0.08},$ where $T_1$
is temperature in units of 10 K. At this rate, turning $30\%$ of the
gas into stars takes $5-6$ dynamical times for a 1000 $M_{\odot}$
clump, with a very weak dependence on temperature, mass, or surface
density.

KM05 modeled clouds that were not
centrally concentrated, so we can improve this estimate by considering
density variation. In turbulent media that are not centrally
condensed, the velocity dispersion increases with length scale $\ell$
as $\sigma\propto \ell^{1/2}$ regardless of position in the medium,
but the tidal field of a clump may introduce a dependence on the
distance $r$ from the clump center as well.  The expected variation is
$\sigma\propto r^{1-\krho/2}$, and observations show $\krho\simeq 1.5$
(Mueller et al. 2002). On length scales $\ell \ll r$, the tidal field
of the clump is negligible and we should find $\sigma\propto
\ell^{1/2}$, as for a uniform medium, while for $\ell \gg r$ the tidal
field will dominate. Since the size scale of a star-forming core is
much less than $r$ over the vast majority of a star-forming clump, we
can approximate this behavior by $\sigma \approx
\sigma_{\rm cl} r^{(1-\krho)/2} R^{\krho/2-1} \ell^{1/2}$ for $\ell
\ll r$. Thus,
the star formation rate varies within the clump as $\mbox{SFR}_{\rm
ff} = \mbox{SFR}_{\rm ff-cl} r^{0.32\,(\krho/2-1)}
R^{0.32\,(1-\krho/2)}$. Similarly, the free-fall time as a function
of distance from the clump center is $\tff = t_{\rm ff-cl}
[3/(3-\krho)]^{1/2} (r/\rcl)^{\krho/2}$. Integrating over radial
shells, the total star formation rate is
\begin{equation}
\dot{M}_* = \int_0^R 4\pi r^2 \rho \frac{\mbox{SFR}_{\rm ff}}{\tff} dr
= \left[\frac{(3-\krho)^{3/2}}{2.3 \,(2-\krho)}\right]
M \frac{\mbox{SFR}_{\rm ff-cl}}{t_{\rm ff-cl}}.
\end{equation}
Thus, central condensation increases the star
formation rate by a factor of $(3-\krho)^{3/2} / [2.3 (2-\krho)]$
relative to a uniform medium. For our fiducial
$\krho=1.5$, this means that the time required to reach $30\%$ star
formation efficiency is reduced by a factor of 1.6 relative to our
previous estimate, giving $3-4$ dynamical times. 

\section{Discussion}

We have presented observational and theoretical evidence that rich
star clusters require at least several dynamical timescales to form.
This is significant, because it implies that star clusters cannot form
by a process of freely decaying turbulence leading to global collapse,
which could not possibly take so long. Instead, something must impede
or entirely prevent global collapse, so that rich star clusters are in
approximate equilibrium during their assembly. For the ONC,
$\rho$~Oph, and IC~348, where we can reasonably estimate both
formation and dynamical times, formation typically requires $\gtsim
3-4$ dynamical times, consistent with theoretical predictions for the
time required to turn $\sim 30\%$ of the gas into stars. The central
regions have much smaller dynamical times than the cluster average, so
we predict that in the center a larger fraction of the gas will form
stars.

Another implication of this work is that massive star feedback
may not be as effective as once assumed in dispersing gas in
young clusters. This is consistent with observations that
massive stars are not always the last to form in their clusters
(e.g. Eisenhauer et al. 1998; Hoogerwerf et al. 2001), and theoretical
work showing that clumpiness greatly inhibits gas dispersal (Tan \&
McKee 2004; Dale et al. 2005). Tan \& McKee find that, in a clump like
the proto-ONC, if $\sim 3\%$ of the mass forms stars per dynamical
time, feedback requires $\sim$ 2~Myr (about 3~$t_{\rm dyn}$) to
disperse the gas. Dynamical ejection of massive stars, as observed in
the ONC (Hoogerwerf et al. 2001; Tan 2004), would increase this time.

Long formation times are also important for mass segregation. For
example, a 3~Myr formation time for the ONC corresponds to about 8
diameter crossing times at the half mass radius of 0.5~pc, which is
the unit of time used in the study of Bonnell \& Davies (1998). If the
30 most massive stars are born at the half-mass radius
then after 3~Myr the median location of the 6 most massive stars
migrates to only $0.075$ pc, suggesting that some of the observed mass
segregation (Hillenbrand \& Hartmann 1998) could be dynamical rather
than primordial. Gas drag will likely enhance the segregation beyond
the purely N-body effects explored by Bonnell \& Davies.

Although undriven supersonic turbulence decays in $\sim 1$ dynamical
time (Stone, Ostriker, \& Gammie 1998), the observed turbulence in
molecular clumps does not damp so quickly. Driving by protostellar
feedback is a possible explanation. A virialized clump that radiates
away half its kinetic energy per dynamical time has a luminosity
$L_{\rm diss} \approx 4 M_3^{5/4} \Sigma_0^{5/4} \:L_\odot$, but the
IMF-averaged energy release associated with accretion is $L_{\rm
acc}\approx 2000 M_3 \Sigma_0^{3/4} (\eta/10)^{-1}\;L_{\odot}$, where
$\eta$ is the number of dynamical times required to transform $50\%$
of the gas into stars. Outflows should eject about half this energy
back into the clump (Shu et al. 2000), so even if only $\sim 1\%$ of
this goes into driving turbulence, that is sufficient to offset the
decay. Recent observations that find outflows inject enough energy to
maintain turbulence (Williams, Plambeck, \& Heyer 2003; Quillen et al. 2005) support this idea, as do the
numerical simulations of Li \& Nakamura (2006).  However, this work is
preliminary and has not yet shown that feedback can
maintain turbulence over a cluster lifetime of ~4 $t_{\rm dyn}$ that
observations seem to require.




\acknowledgements We thank B. Elmegreen, F. Palla, S. Stahler \&
S. Tremaine for discussions. We acknowledge support from: an ETH
Z\"urich Zwicky Fellowship (JCT); NASA via Hubble Fellowship grant
\#HSF-HF-01186 from STScI, operated by AURA
for NASA under contract NAS 5-26555 (MRK); NSF grant AST-0098365
(CFM); and NASA ATP grant NAG 5-12042 (CFM).


\begin{references}
\reference{} Bate, M. R., Bonnell, I. A., \& Bromm, V. 2003, \mnras, 339, 577
\reference{} Bertoldi, F., \& McKee, C. F. 1992, \apj, 395, 140
\reference{} Bonnell, I. A., \& Bate, M. R. 2002, \mnras, 336, 659
\reference{} Bonnell, I. A., \& Davies, M. B. 1998, \mnras, 295, 691
\reference{} Binney, J., \& Tremaine, S. 1987, {\it Galactic Dynamics}, Princeton
\reference{} Burningham, B., Naylor, T., Littlefair, S.~P., \& Jeffries, R.~D. 2005, \mnras, 363, 1389
\reference{} Dale, J. E., Bonnell, I. A., Clarke, C. J., \& Bate,
M. R. 2005, \mnras, 358, 291
\reference{} Egan, M. P., {Shipman}, R. F., {Price}, S. D., {Carey}, S. D., {Clark}, F. O., {Cohen}, M. 1998, \apj, 494, L199
\reference{} Eisenhauer, F. Quirrenbach, A., Zinnecker, H., \& Genzel, R. 1998, \apj, 498, 278
\reference{} Elmegreen, B. G. 2000, \apj, 530, 277
\reference{} Forbes, D. 1996, \aj, 112, 1073
\reference{} Goodwin, S. P., \& Whitworth, A. P. 2004, \aap, 413. 929
\reference{} Hartmann, L. 2003, \apj, 585, 398
\reference{} Hartmann, L., Ballesteros-Paredes, J., \& Bergin, E. 2001, \apj, 562, 852 
\reference{} Herbig, G. H. 2005, \apj, 497, 736
\reference{} Hillenbrand, L. A. 1997, \aj, 113, 1733
\reference{} Hillenbrand, L. A., \& Hartmann, L. W. 1998, \apj, 492, 540
\reference{} Hodapp, K-W, \& Deane, J. 1993, \apjs, 88, 119
\reference{} Hoogerwerf, R., de Bruijne, J., \& de Zeeuw, P.T. 2001, \aap, 365,49
\reference{} Krumholz, M. R., \& McKee, C. F. 2005, \apj, 630, 250
\reference{} Lada, E. A., \& Lada, C. J. 1995, \aj, 109, 1682
\reference{} Li, Z-Y., \& Nakamura, F. 2006, \apj, sub, (astro-ph/0512278)
\reference{} Loren, R. B. 1989, \apj, 338, 902
\reference{} McKee, C. F., \& Tan, J. C. 2003, \apj, 585, 850
\reference{} Mueller, K. E., Shirley, Y. L., Evans, N. J. II, Jacobson, H. R. 2002, \apjs, 143, 469
\reference{} Najita, J. R., \& Shu, F. H. 1994, \apj, 429, 808
\reference{} Palla, F., Randich, S., Flaccomio, E., Pallavicini, R. 2005, \apj, 626, L49
\reference{} Palla, F., \& Stahler, S. W. 1999, \apj, 525, 722 (PS99)
\reference{} Palla, F., \& Stahler, S. W. 2000, \apj, 255, 270
\reference{} Quillen, A. C. et al. 2005, \apj, 632, 941
\reference{} Scally, A., \& Clarke, C. 2002, \mnras, 334, 156
\reference{} Schmeja, S. \& Klessen, R. S. 2006, \aap, sub, (astro-ph/0511448)
\reference{} Shirley, Y. L., Evans, N. J. II, Young, K. E., Knez, C., Jaffe, D. T. 2003, \apjs, 149, 375 
\reference{} Shu, F. H., Najita, J., Shang, H., Li, Z.-H. 2000, Protostars \& Planets IV, ed. Mannings, Boss, \& Russell (Tucson:Arizona),789
\reference{} Stahler, S. W. 1988, \apj, 332, 804
\reference{} Stone, J. M., Ostriker, E. C., \& Gammie, C. 1998, \apjl, 508, L99
\reference{} Tan, J. C. 2004, \apj, 607, L47
\reference{} Tan, J. C., \& McKee, C. F. 2002, Hot Star Workshop III: Earliest Stages of Massive Star Birth, ed. Crowther, ASP, 267, 267 
\reference{} Tan, J. C., \& McKee, C. F. 2004, Formation \& Evolution of Young Massive Clusters, eds. Lamers, Nota \& Smith, ASP, 322, 263
\reference{} Tassis, K., \& Mouschovias, T. Ch. 2004, \apj, 616, 283
\reference{} Williams, J. P., Plambeck, R. L., \& Heyer, M. H. 2003, \apj, 591, 1025
\end{references}
\end{document}